\documentstyle[sprocl,epsf]{article}

\def\Journal#1#2#3#4{{#1} {\bf #2}, #3 (#4)}

\def\NPB{{\em Nucl. Phys.} B}

\def\be{\begin{equation}}
\def\ee{\end{equation}}
\def\bea{\begin{eqnarray}}
\def\eea{\end{eqnarray}}

\begin{document}

\title{COLOUR CONFINEMENT IN THE \\
LATTICE LANDAU GAUGE QCD SIMULATION}

\author{Sadataka FURUI}

\address{School of Science and Engineering, Teikyo University,
\\E-mail: furui@liberal.umb.teikyo-u.ac.jp} 

\author{Hideo NAKAJIMA}

\address{Department of Information Science, Utsunomiya University,
\\E-mail: nakajima@is.utsunomiya-u.ac.jp}


\maketitle\abstracts{ The colour confinement criterion proposed by Kugo and 
Ojima is tested in the lattice Landau gauge QCD simulation. The renormalization
effects are studied by measuring the gluon propagator, ghost propagator,
three gluon vertex and the ghost-antighost-gluon vertex. 
The running coupling $\alpha_s$ from ghost-antighost-gluon vertex in the
infrared region yields $\Lambda_{\widetilde{MOM}}\simeq 1 GeV$, consistent
to that from three gluon vertex in high momentum region. }

\section{The confinement signal and the definition of the gauge field}
Two decades ago, Gribov pointed out a possible mechanism of colour confinement
in Coulomb gauge or Landau gauge QCD via infrared divergence of the 
Faddeev-Popov ghost propagator\cite{Gv}. 
At nearly the same time, Kugo and Ojima proposed a criterion for the absence
of coloured massless asymptoptic states in Landau gauge QCD using the BRST
symmetry\cite{KO}.

We study the confinement signal in the gluon propagator, the 
ghost propagator and the Kugo-Ojima parameter, and their dependence on
the gauge field.
Usually, the gauge field on lattice
$A_\mu(x)$ is defined from the link variable $U_\mu(x)$ as 
$A_{x,\mu}={1\over 2}(U_{x,\mu}-U_{x,\mu}^{\dag})|_{traceless\ part}$, which we call $U$-linear version. A more natural definition\cite{NF} is
$U_{x,\mu}=\exp{A_{x,\mu}},\ \ \ A_{x,\mu}^{\dag}=-A_{x,\mu}$,
which we call $\log U$ version. 

We observe that the Kugo-Ojima parameter
$u^{ab}(0)$ at $\beta=5.5$ is about -0.7 in the $\log U$ and about -0.6 in the
$U$-linear version.
\begin{figure}[htb]
\begin{minipage}[b]{0.47\linewidth} 
\begin{center}
\epsfysize=80pt\epsfbox{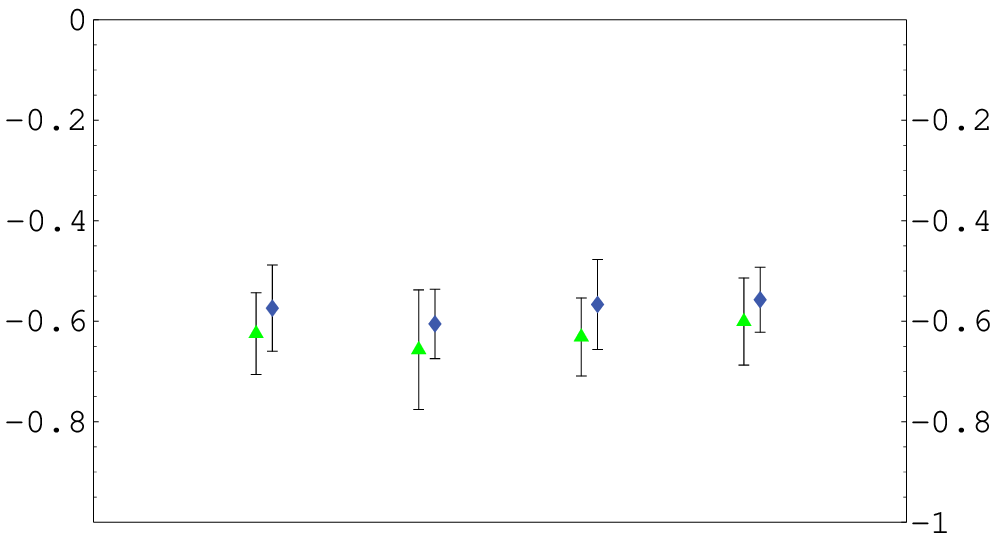}
\caption{The dependence of the Kugo-Ojima parameter measured along x,y,z and t axis resp. on the definition
of the gauge field. $\beta=6.0, 16^4$.
The triangles are $\log U$, and diamonds are $U-$ linear version.}
\end{center}
\end{minipage}
\hfil
\begin{minipage}[b]{0.47\linewidth} 
\begin{center}
\epsfysize=100pt\epsfbox{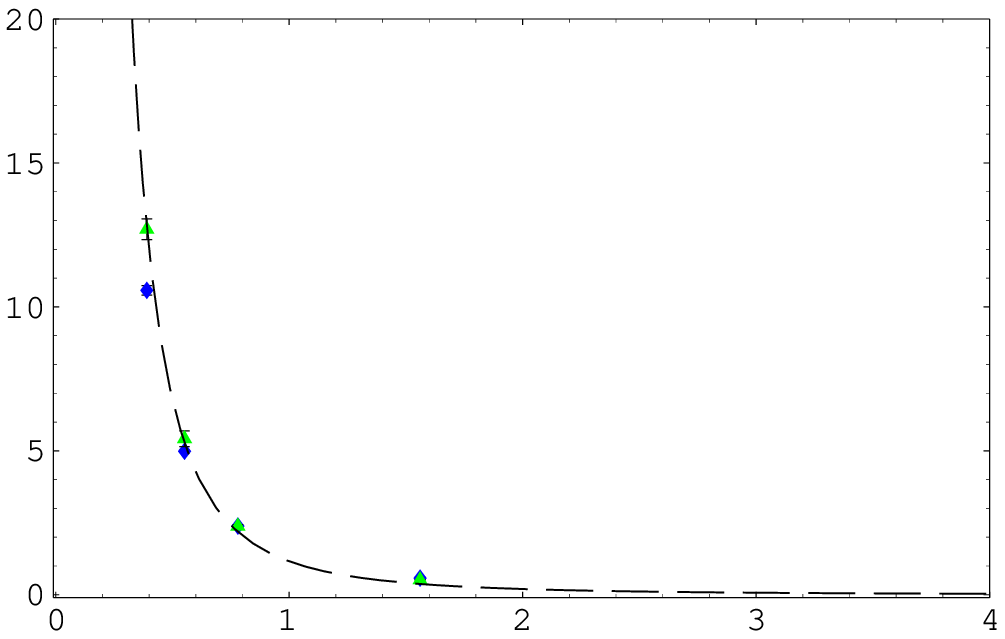}
\caption{The ghost propagator as a function of the lattice momentum. $\beta=6.0$, $16^4$. Triangles are $\log U$ definition and diamonds are $U$-linear definition. The dashed curve is $1.162/ p^{2.545}$}
\end{center}
\end{minipage}
\end{figure}

If the configuration is in the core region\cite{Zw},
the tensor $G_{\mu\nu}^{ab}=tr(\bar\lambda^a D_\mu{1\over -\partial D}
(-D_\nu)\lambda^b)_{xy}$ divided by $N^2-1$, where $N$ is the number of colours,  is expected to approach
a function $E(U)$ defined by the optimizing function\cite{Zw}. 
They are $E(U)=\sum_l{1\over N}{\rm Re}\ {\rm tr} U_l$ in $U$-linear version, 
and
${{1\over N^2-1}\sum_{l,a}}{\rm tr}\left ({\lambda^a}^\dag S({\cal A}_l)\lambda^a\right)$, where ${\cal A}_l=adj_{A_l}$ and $S(x)={{x/2\over {\rm th}(x/2)}}$,
in $\log U$ version. 

In the table below $e_1$
and $e_2$ stand for $e=\langle {E(U)\over  V}\rangle$, in our $16^4$ lattice simulation of the $U$-linear
and the $\log U$ version of the gauge fields, respectively. We define the
horizon function $h=\langle {H(U)\over V}\rangle/(3 (N^2-1))=c-e/d$.  In general we expect $h<0$
and in the continuum limit, we expect $h=0$ when the configuration is in the 
core region.

\begin{table}[htb]
\caption{$\beta$ dependence of the Kugo-Ojima parameter $c$, 
 trace $e$ divided by the dimension $d$, and $h=c-e/d$.
The suffix 1 corresponds to the $u$-linear and 2 corresponds to  the $\log U$
version. Data are those of $16^4$, except  $\beta=5.5$ $U$-linear data,
which are those of $8^4$. } 
\vskip 0.5 true cm
\begin{tabular*}{\textwidth}{@{}l@{\extracolsep{\fill}}c|ccc|cccc}
    & $\beta$    & $c_1$  & $e_1/d$  & $h$&$c_2$& $e_2/d$& $h$ &\\
\hline
  &5.5  & 0.570(58)& 0.780(3) & -0.21 &0.712(18)  &0.908(1) &  -0.20 \\
 &6.0   &0.576(79)& 0.860(1) & -0.28 & 0.628(94)  & 0.943(1)  & -0.32 \\
\hline
\end{tabular*}
\end{table}

\begin{figure}[htb]
\begin{minipage}[b]{0.47\linewidth} 
\begin{center}
\epsfysize=100pt\epsfbox{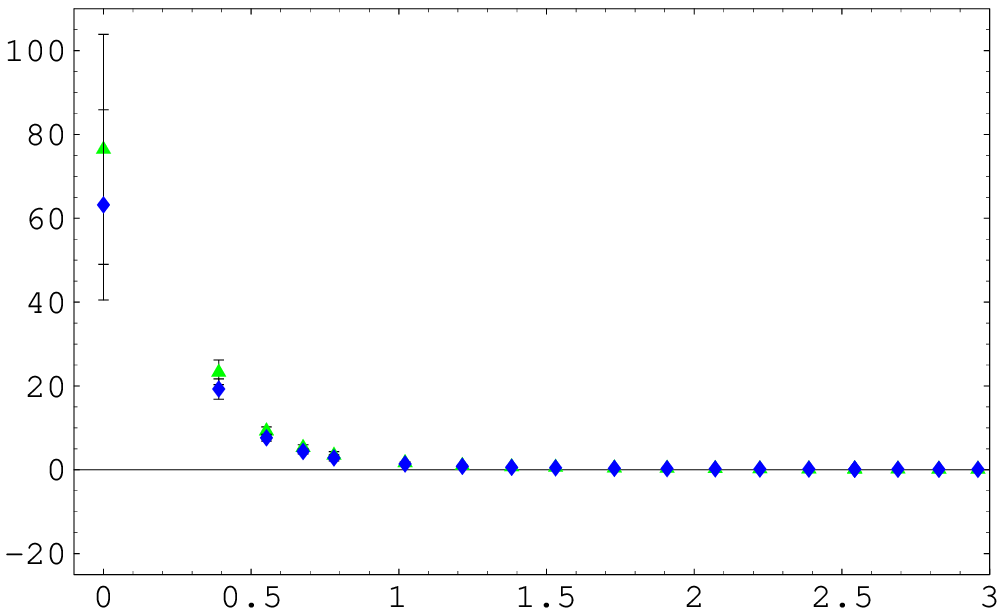}
\caption{The gluon propagator as a function of the lattice momentum. $\beta=6.0$, $16^4$. Triangles are $\log U$ version and diamonds are $U$-linear version. 
50 samples.}
\label{gl3.2}
\end{center}
\end{minipage}
\hfil
\begin{minipage}[b]{0.47\linewidth} 
\begin{center}
\epsfysize=100pt\epsfbox{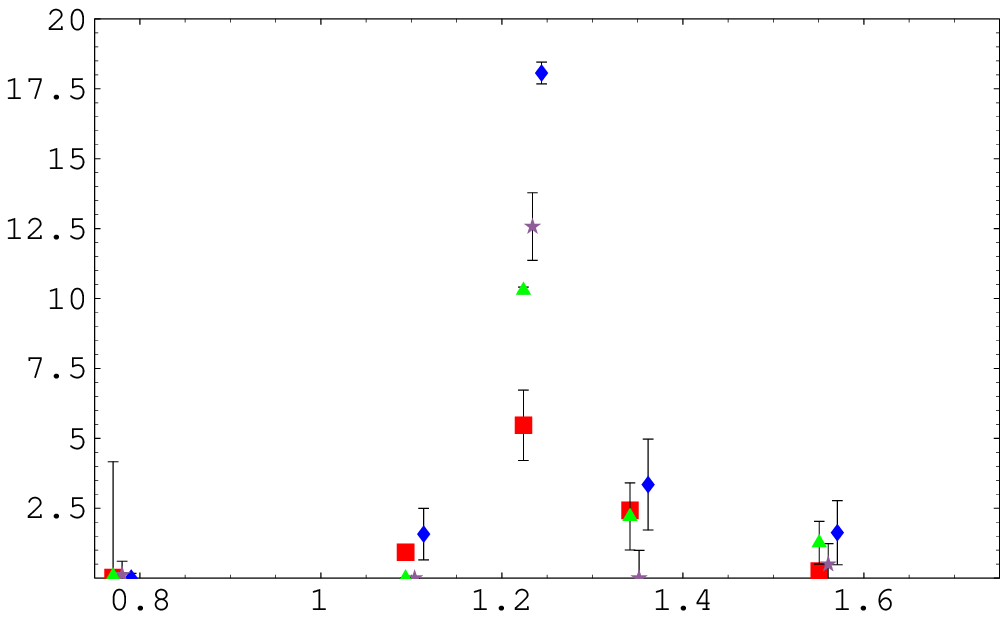}
\caption{The QCD running coupling constant as a function of momentum
$\mu$. Triangles and diamonds are $g^2/4\pi$, the same as Fig\ref{gl3.2}, stars and boxes are $g^2/4\pi$ and $\tilde g^2/4\pi$, respectively of $\log U$ smeared version.}
\label{alp}
\end{center}
\end{minipage}
\hfil
\end{figure}
The gluon propagator is infrared finite and the absolute value in $\log U$ 
version is about 20\% larger than that in $U$-linear version
The corresponding difference in the ghost propagator is about 10\%.

\section{The QCD running coupling}

The QCD running coupling $\alpha(\mu)=g^2/4\pi$ can be measured from the
three gluon vertices\cite{Orsay} as,
$g(\mu^2)={{G_A}^{(3)}({p_i}^2,{p_f}^2,{p_c}^2) Z_3^{3/2}(\mu^2)
\over {G_A}^{(2)}({p_i}^2){G_A}^{(2)}({p_f}^2){G_A}^{(2)}({p_c}^2)}$.
Since the lattice data of ${G_A}^{(2)}$ is infrared finite, in contrast to the
 conjecture that the gluon propagator is infrared 
vanishing\cite{Zw}, $g(\mu^2)$ 
decreases as $\mu$ decreases as $1.5\mu$.
 This behaviour does not agree with the
results of Dyson-Schwinger approach\cite{SHA}, which suggest that $\alpha_s(\mu)$ monotonically increases to a finite constant as $\mu$ goes to 0. 

The running coupling can be obtained also from 
the ghost-antighost-gluon coupling 
$\tilde g(\mu^2)={{G_g}^{(3)}({p_i}^2,{p_f}^2,{p_c}^2) Z_3^{1/2}(\mu^2)
\tilde Z_3(\mu^2)\over {G_g}^{(2)}({p_i}^2){G_g}^{(2)}({p_f}^2){G_A}^{(2)}({p_c}^2)}$.
The preliminary results of  $g^2/4\pi$ and $\tilde g^2/4\pi$
at symmetric momentum points $\langle p_\mu\rangle={2\over a} \sin ({n_\mu \pi/L})$, $\sum_\mu n_\mu^2=10$ have a peak which depend on definition of the $A_\mu$ but at other momentum points $g^2/4\pi$ and $\tilde g^2/4\pi$ are not so different.
When the physical scale is fixed by $a^{-1}=1.91\pm 0.1 GeV$, 
$\Lambda_{\widetilde{MOM}}$ calculated from the $\tilde g^2/4\pi$ at $\sum_\mu n_\mu^2=8$ and $16$ are about $1 GeV$, which are consistent to that obtained from the three gluon vertex in high momentum region\cite{Orsay}. 

S.F. thanks Prof. Reinhardt and Prof. Alkofer for helpful discussion and 
hospitality in Tuebingen in August 2000.
This work is supported by JSPS Grant-in-aid No.11640251 and the KEK supercomputer project No.00-57.

\end{document}